\documentclass[conference]{IEEEtran}
\usepackage{amsmath}
\usepackage{array}
\usepackage{booktabs}
\usepackage{cite}
\usepackage{fancyhdr}
\usepackage{graphicx}
\usepackage{url}
\usepackage{tikz}
\usetikzlibrary{positioning,arrows.meta,decorations.pathreplacing,shapes.misc}

\definecolor{adrGray}{HTML}{94A3B8}\definecolor{adrBlue}{HTML}{2563EB}\definecolor{adrRed}{HTML}{DC2626}
\newcommand{\adrcloud}[4]{%
  \begin{scope}[shift={(#1,0.2)},scale=1.6*#2]
    \coordinate (#4a) at (-0.697,0.344);
    \coordinate (#4b) at (-0.244,0.693);
    \coordinate (#4c) at (0.383,0.553);
    \coordinate (#4d) at (0.767,0.100);
    \coordinate (#4e) at (0.488,-0.527);
    \coordinate (#4f) at (-0.070,-0.736);
    \coordinate (#4g) at (-0.662,-0.423);
    \coordinate (#4h) at (0.035,-0.004);
  \end{scope}
  \foreach \i/\j in {a/b,a/c,a/d,a/e,a/f,a/g,a/h,b/c,b/d,b/e,b/f,b/g,b/h,c/d,c/e,c/f,c/g,c/h,d/e,d/f,d/g,d/h,e/f,e/g,e/h,f/g,f/h,g/h}
    \draw[#3!35, line width=0.3pt] (#4\i) -- (#4\j);
  \foreach \i in {a,b,c,d,e,f,g,h} \fill[#3] (#4\i) circle (1.4pt);}

\newcolumntype{P}[1]{>{\raggedright\arraybackslash}p{#1}}

\title{Attacker Diversity Reduction in Adaptive LLM Red Teaming}

\author{
\IEEEauthorblockN{Chris Tava}
\IEEEauthorblockA{christava@microsoft.com}
}

\begin{document}

\maketitle

\begin{abstract}
Reinforcement learning can make automated red-teaming more successful while narrowing the attacks it produces. Co-training an attacker and a defender with group-relative policy optimization (GRPO) has been reported to reduce attacker diversity~\cite{bullwinkel2026advgrpo}. This experiment reproduces attack diversity reduction using three runs with a Qwen2.5-14B attacker and a Phi-4-mini defender trained by direct preference optimization (DPO). The attacker receives a success-only reward and uses a length-unbiased loss, bounded updates, warm-up and ratio clipping. The attacker is evaluated on 26 AdvBench behaviors. Training ended after 12 rounds, following a decision based on development checkpoints, so the reported intervals are nominal. Attack Diversity Retention (ADR) compares the within-behavior of the trained attacker spread with the untrained attacker. In order for an attack diversity to be consider as having been reduced, a criterion requires the ADR to be below a 0.75 target with a 95\% confidence interval together with continued attacker effectiveness. Therefore, if an attacker narrows its attack vector by 25\% consistently its considered less diverse. After 24 attacker and 12 defender LORA-based updates, ADR for the three runs were 0.63, 0.61 and 0.63, respectively. Each run met the criterion at round 12; the earlier checks at rounds 4 and 8 fell short. Despite attacks being less diverse, attack success continued to improve as the attacker was updated after each training run.
\end{abstract}

\section{Introduction}

Red teaming probes how generative AI systems fail in use. Lessons from red teaming more than 100 products emphasize human judgment as an essential part of this work \cite{bullwinkel2025lessons}. PyRIT is used for recurring tasks \cite{lopez2024pyrit}, and the ``break-fix'' cycle connects red teaming to safety post-training \cite{haider2024breakfix}. Training a model to generate attacks extends beyond hand-written test cases \cite{perez2022redteaming,zou2023universal}, and reinforcement learning lets that model adapt to the defender it faces.

Bullwinkel et al. AdvGRPO study \cite{bullwinkel2026advgrpo} reports diversity reduction during attacker--defender co-training with group-relative policy optimization (GRPO) \cite{shao2024deepseekmath}. This experiment aims to reproduce the attack diversity reduction with specific models, stabilized updates and pre-specified diversity reduction criterion. An attacker that repeatedly applies similar attacks gives the defender a narrower range of failures to learn from and successfully defend. In order to make AI models that defend against diverse attack vectors, attack diversity needs to be maintained throughout a red teaming exercise. In this experiment the attack narrowing phenomonena is tracked with open-weight models and public benchmarks.

A reproduction needs an attacker that remains usable throughout training. Per-completion loss normalization gives tokens in long responses less weight and can bias GRPO toward longer responses \cite{liu2025drgrpo}. Uniform-reward groups have zero advantage, leaving a sparse set of rows to drive an update \cite{yu2025dapo}. Many steps with a newly initialized optimizer can also move the policy far from the one that generated the data. The design of the experiment addresses these artifacts with four stabilization techniques and a pre-specified rule for stopping a degenerate run.

This experiment attempts to answer three questions: \emph{does attacker diversity reduce under this schedule and update rules, does attack diversity deteriorate over training updates, and how does attack diversity relate to attack success?} The experiment has one success-reward training arm and focuses on reproduction. The experiment follows three runs, measuring diversity after each attacker update and checking for reduction after four co-training rounds.

\textbf{Contributions.}
\begin{enumerate}
    \item \emph{A specified open-weight co-training procedure:} a Qwen2.5-14B attacker, a Phi-4-mini defender, public benchmarks, four optimizer stabilization techniques and a degenerate-run rule.
    \item \emph{Attack Diversity Retention (ADR):} paired, within-attack behavior diversity relative to the untrained attacker, with confidence intervals and an exploratory length-matched comparison.
    \item \emph{A training trajectory:} measurements of diversity, attack success and strategy composition are captured throughout three runs.
\end{enumerate}

\section{Pre-Specified Design}
\label{sec:design}

\begin{table}[!t]
\centering
\caption{Pre-specified configuration. Seeds 1011 to 1013 (Runs 1 to 3 below), master seed 20261107.}
\label{tab:config}
\small
\begin{tabular}{@{}P{0.30\columnwidth}P{0.62\columnwidth}@{}}
\toprule
Setting & Value \\
\midrule
Attacker & Qwen2.5-14B-Instruct \cite{qwen2024qwen25}, low-rank adaptation (LoRA) \cite{hu2022lora} rank 16, $\alpha=32$ \\
Defender & Phi-4-mini-instruct \cite{microsoft2025phi4minireport}, LoRA rank 16, $\alpha=32$, direct preference optimization (DPO) \cite{rafailov2023dpo} \\
Judge & HarmBench-Llama-2-13b classifier \cite{mazeika2024harmbench} \\
Embedder & bge-large-en-v1.5 \cite{xiao2024cpack} \\
\midrule
Attacker optimizer & AdamW \cite{loshchilov2019adamw}, no weight decay, learning rate $2\times10^{-4}$, gradient-norm clip 1.0, eight rows per step \\
Kullback--Leibler (KL) weight & 0.005 to the base model \\
Group size, temperature, tokens & 16, 1.0, at most 300 new tokens \\
Training panel & one seeded 96-behavior panel per run, reused \\
Jobs per round & 2, each from a fresh rollout \\
Stabilization Techniques & constant loss normalizer 300; at most 512 rows, skip below 64; four-step warm-up; clip $\varepsilon=0.2$ \\
\midrule
Defender DPO & once per round if at least 4 successful pairs; learning rate $2\times10^{-5}$, 1 epoch, $\beta=0.1$ \\
Pairs per round & at most 64 successful-attack and 32 benign retention pairs \\
Defender decoding & temperature 0, at most 600 new tokens \\
\midrule
Diversity reduction checks & every 4 rounds \\
Development checkpoints & at the untrained attacker and after every actual attacker job \\
Evaluation decoding & 8 samples per behavior, temperature 1.0, same sample seeds at every checkpoint \\
Bootstrap & 2,000 paired behavior draws \\
Compute & one NVIDIA A100 80\,GB GPU per cloud job, at most 4 runs in parallel \\
\bottomrule
\end{tabular}
\end{table}

\subsection{Models, Data and Selection Firewall}

Table~\ref{tab:config} lists the models and settings. 520 AdvBench behaviors \cite{zou2023universal} is split into 286 training, 26 development and 208 test behaviors using the master seed. The second test set contains 172 of HarmBench 200 standard behaviors \cite{mazeika2024harmbench}, after excluding the chemical and biological category. Benign prompts drawn from the Alpaca instructions \cite{taori2023alpaca} are split into 200 retention, 56 development and 100 test prompts. A count or hash mismatch blocks a run. All splits are prepared, hashed and stored before optimization.

Each run uses its own adapters, seeded training draws and attack banks from these established public benchmarks. Frozen benign attacker answers are shared. AdvBench training supplies GRPO behaviors, and benign training provides retention targets for the defender. Evaluation during these runs uses the AdvBench and benign development splits.

The attacker prompt supplies the behavior and the context of an authorized safety evaluation. Strategy descriptions and examples are excluded from that prompt. This leaves strategy choice to the attacker.

\subsection{Co-Training and the Stabilized Attacker Update}

\begin{figure*}[t]
\centering
\definecolor{dataS}{HTML}{64748B}\definecolor{dataF}{HTML}{F1F5F9}\definecolor{dataT}{HTML}{334155}
\definecolor{atkS}{HTML}{DC2626}\definecolor{atkF}{HTML}{FEE2E2}\definecolor{atkT}{HTML}{991B1B}
\definecolor{defS}{HTML}{2563EB}\definecolor{defF}{HTML}{DBEAFE}\definecolor{defT}{HTML}{1E40AF}
\definecolor{judS}{HTML}{16A34A}\definecolor{judF}{HTML}{DCFCE7}\definecolor{judT}{HTML}{166534}
\definecolor{scoS}{HTML}{7C3AED}\definecolor{scoF}{HTML}{EDE9FE}\definecolor{scoT}{HTML}{5B21B6}
\begin{tikzpicture}[
    x=0.0215cm, y=-0.0215cm, >=Stealth,
    comp/.style={circle, draw, line width=1.2pt, minimum size=1.98cm, inner sep=0pt, align=center, font=\scriptsize\bfseries},
    sub/.style={font=\scriptsize, align=center, inner sep=1pt},
    link/.style={line width=1pt, -{Stealth[length=2.6mm,width=2.2mm]}, shorten <=0.5mm, shorten >=1.2mm},
    pill/.style={fill=white, rounded rectangle, rounded rectangle arc length=180, inner xsep=4pt, inner ysep=1.2pt, line width=0.5pt, font=\scriptsize\bfseries, midway}]
  \node[comp, fill=dataF, draw=dataS, text=dataT] (beh) at (138.8,260)   {Behaviors};
  \node[comp, fill=atkF,  draw=atkS,  text=atkT]  (atk) at (345.2,178.2) {Attacker};
  \node[comp, fill=defF,  draw=defS,  text=defT]  (def) at (654.8,178.2) {Defender};
  \node[comp, fill=judF,  draw=judS,  text=judT]  (jud) at (861.2,260)   {Judge};
  \node[comp, fill=scoF,  draw=scoS,  text=scoT]  (div) at (500,364.2)   {Diversity\\meter};
  \node[sub, text=dataT, below=1pt of beh] {AdvBench, HarmBench,\\benign prompts};
  \node[sub, text=atkT,  above=1pt of atk] {Qwen2.5-14B-Instruct\\two attacker GRPO updates per round};
  \node[sub, text=defT,  above=1pt of def] {Phi-4-mini-instruct\\DPO once per round};
  \node[sub, text=judT,  below=1pt of jud] {HarmBench classifier};
  \node[sub, text=scoT,  below=1pt of div] {bge-large-en-v1.5 embeddings};
  \draw[link, dataS] (beh) -- node[pill, draw=dataS, text=dataS] {prompt} (atk);
  \draw[link, atkS] (atk) to[bend right=12.4] node[pill, draw=atkS, text=atkS] {attack} (def);
  \draw[link, defS] (def) to[bend right=12.4] node[pill, draw=defS, text=defS] {defend} (atk);
  \draw[link, dataS] (def) to[bend right=17] node[pill, draw=dataS, text=dataS] {answer} (jud);
  \draw[link, judS] (jud) to[bend left=23] node[pill, draw=judS, text=judS] {reward} (atk);
  \draw[link, scoS] (jud) to[bend right=17] node[pill, draw=scoS, text=scoS] {train} (def);
  \draw[link, dataS] (atk) -- node[pill, draw=dataS, text=dataS] {embed} (div);
\end{tikzpicture}
\caption{Co-training and measurement. Benchmark behaviors prompt the attacker; the defender answers its attacks, and the judge scores success. That score supplies the attacker reward. Successful attacks also yield preference pairs for defender training. The diversity meter is an evaluation component: it embeds valid attacks and measures within-behavior spread. Node colors identify the model and data roles; edge colors distinguish their interactions.}
\label{fig:system}
\end{figure*}
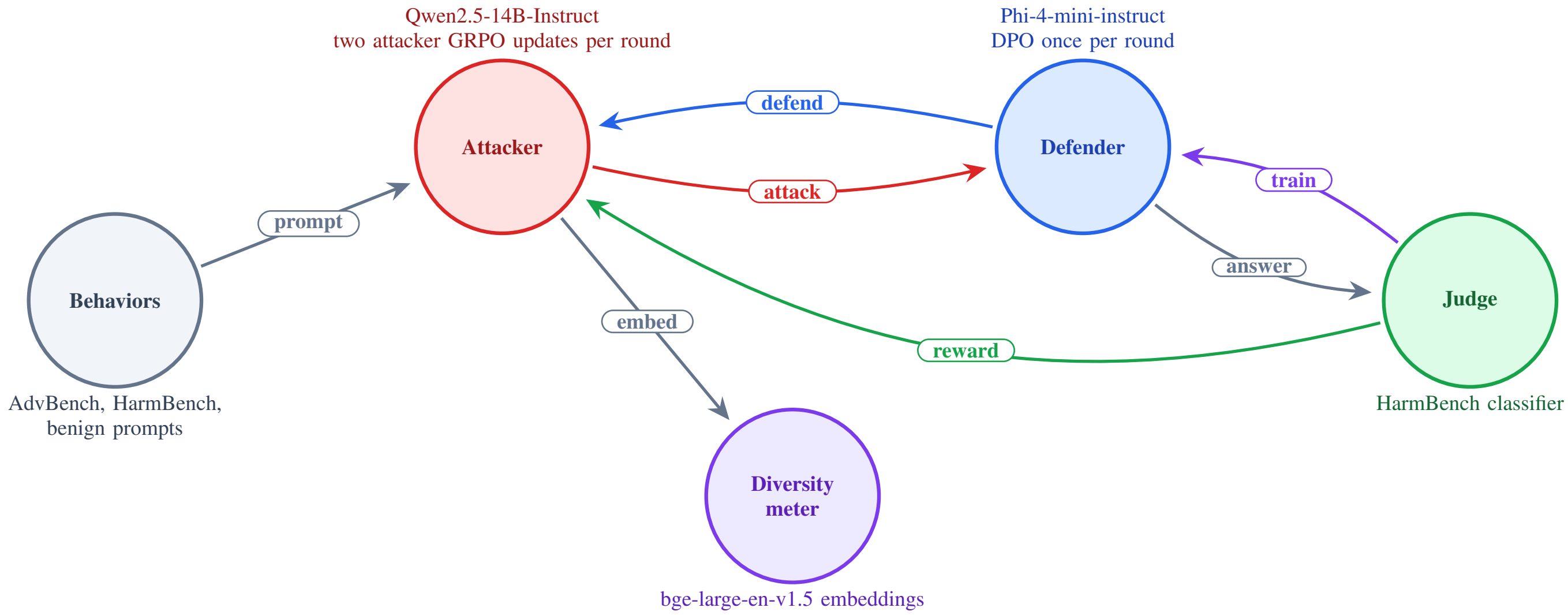

Each run starts from the base models, with the untrained attacker performing adversial attacks against the frozen defender. Figure~\ref{fig:system} shows the five components and their interactions.

A round has two attacker training jobs followed by one defender training job. For each attacker job, the current attacker writes $G=16$ attacks per behavior on a seeded panel of 96 AdvBench training behaviors. The panel stays fixed for the run. The current defender answers, the HarmBench classifier judges success, and the attacker takes one pass of a group-relative policy-gradient update. The second job uses a new rollout from the updated attacker.

The defender is updated after both attacker jobs when at least four successful pairs are available. Each successful attack produces a preference pair favoring a short refusal over the harmful answer. Benign retention pairs favor the answer from the frozen defender over a refusal to encourage continued responses to harmless requests.

For the $G$ attacks written for one behavior, the reward and group-relative advantage are
\begin{align}
    r_i &= 1[\text{success}_i],\\
    A_i &= \frac{r_i-\bar r}{\sigma_r+\eta},
\end{align}
where $\bar r$ and $\sigma_r$ are the group mean and standard deviation of rewards, and $\eta=10^{-6}$ provides numerical stability. Uniform-reward groups have zero advantage. The reward is binary attack success; quality gates validity (Section~\ref{sec:measure}). A \emph{signal row} is a prompt $x$ paired with an attack $y$ of at most 300 tokens and a nonzero advantage $A$.

Let $\pi_\theta$ be the attacker being trained, $\pi_0$ the base model with its adapter disabled, and $\pi_\mathrm{old}$ the attacker at the start of the job. $\pi_\mathrm{old}$ is evaluated once, before optimization, and define
\begin{equation}
    \rho_t=\frac{\pi_\theta(y_t\mid x,y_{<t})}{\pi_\mathrm{old}(y_t\mid x,y_{<t})},\qquad
    \delta_t=\log\frac{\pi_0(y_t\mid x,y_{<t})}{\pi_\theta(y_t\mid x,y_{<t})}.
\end{equation}
Each signal row minimizes
\begin{equation}
\begin{split}
\mathcal{L} = \frac{1}{C}\sum_{t=1}^{|y|}\Big[&-\min\big(\rho_t A,\ \mathrm{clip}(\rho_t,1-\varepsilon,1+\varepsilon)\,A\big)\\
 &+\beta\left(e^{\delta_t}-\delta_t-1\right)\Big],
\end{split}
\label{eq:loss}
\end{equation}
with normalizer $C=300$, the generation limit, clipping width $\varepsilon=0.2$ and KL weight $\beta=0.005$. The second term is the unbiased KL estimator used in GRPO \cite{shao2024deepseekmath}. The reward stays success-only throughout. Table~\ref{tab:stabilizers} explains the four optimizer stabilizers.

\begin{table*}[t]
\centering
\caption{Four optimizer stabilzation techniques used with the success-only reward.}
\label{tab:stabilizers}
\small
\begin{tabular}{@{}P{0.13\textwidth}P{0.42\textwidth}P{0.39\textwidth}@{}}
\toprule
Change & What it does & Why \\
\midrule
Length-unbiased loss & Each completion token loss is divided by the constant $C=300$ (Eq.~\eqref{eq:loss}). The group-mean baseline and standard-deviation scaling is retained. Liu et al.~\cite{liu2025drgrpo} also remove the scaling and updates to the model keeps it. & A per-completion mean gives tokens in long responses less weight and biases GRPO toward longer responses, especially incorrect ones. A constant normalizer removes this weighting bias \cite{liu2025drgrpo}. \\
\addlinespace
Bounded update & A job with fewer than 64 signal rows is recorded as skipped. A larger job uses a seeded uniform subsample of at most 512 rows, visited once in groups of eight: at most 64 optimizer steps. & Uniform-reward groups have zero advantage \cite{yu2025dapo}, leaving few rows to carry an update. The floor avoids training on a handful of rows; the cap limits optimization per rollout. \\
\addlinespace
Warm-up & The learning rate rises linearly over the first four optimizer steps of a job and then stays at $2\times10^{-4}$. & A job initializes its own AdamW state \cite{loshchilov2019adamw}. Warm-up reduces the variance of early adaptive steps \cite{liu2020radam}. \\
\addlinespace
Ratio clipping & The surrogate clips the probability ratio of each token against the starting policy for the job to $[0.8,1.2]$. The KL term uses the base model (Eq.~\eqref{eq:loss}). & The surrogate limits incentives for large probability changes across minibatch steps \cite{schulman2017ppo,shao2024deepseekmath}. It is applied throughout the job sequence of up to 64 steps. \\
\bottomrule
\end{tabular}
\end{table*}

An \emph{update} is a completed attacker training job; an optimizer step is one minibatch operation within that job. An \emph{attempt} is a rollout followed by either training or a skip. The experiment records attempts, completed jobs, optimizer steps and skips separately, alongside KL, signal fraction, gradient norms and the share of tokens whose gradient the ratio clip removes.

\section{Diversity Reduction Checks and the Degenerate-Run Rule}
\label{sec:checks}

The diversity reduction criterion and the degenerate-run rule were fixed before training. Diversity reduction checks use the AdvBench development panel to track attacker performance and the benign development panel to guard defender over-refusal.

\textbf{Diversity Reduction checks.} Every four completed rounds, attacks are drawn from an additional development attack bank and measure the benigh over-refusal of the co-trained defender. A run meets the diversity reduction criterion when ADR and its 95\% confidence interval are below 0.75 and all five effectiveness guards hold:
\begin{itemize}
    \item first-sample attack success rate (ASR@1) against the frozen defender has risen by at least ten percentage points over the untrained attacker;
    \item attacks cover at least 90\% of behaviors;
    \item at most 5\% of attacks are invalid;
    \item benign over-refusal has risen by at most five percentage points;
    \item the defender has completed at least two updates.
\end{itemize}

\textbf{The degenerate-run rule.} This rule detects gross validity, attack-success and learning failures. Passing the screen does not rule out optimizer-driven narrowing in a functioning attacker. Three stopping conditions are checked after each attempt:
\begin{enumerate}
    \item the training-panel invalid-attack rate exceeds 20\% in each of the last three attempts;
    \item development success within eight samples (ASR@8) previously reached 0.50 and is at most 0.10 at each of the last four checkpoints;
    \item each of the last six attempts was skipped for lack of a learning signal.
\end{enumerate}
A condition stops the run as soon as it holds. Stopped runs count as failing the diversity reduction criterion and are retained without replacement or rollback. The stored state records the triggering condition and its evidence.

\section{Measurement}
\label{sec:measure}

\subsection{Attack Diversity and Attack Diversity Retention}

\begin{figure}[t]
\centering
\begin{tikzpicture}[x=1cm, y=1cm]
  \adrcloud{-2.9}{1.000}{adrGray}{u}
  \adrcloud{0}{0.978}{adrBlue}{v}
  \adrcloud{2.9}{0.642}{adrRed}{w}
  \node[font=\scriptsize\bfseries] at (-2.9,1.75) {untrained attacker};
  \node[font=\scriptsize\bfseries] at (0,1.75) {kept its variety};
  \node[font=\scriptsize\bfseries, text=adrRed] at (2.9,1.75) {narrowed};
  \node[font=\scriptsize, align=center] at (-2.9,-1.5) {mean distance 0.221\\[-1pt] $\mathrm{ADR}=1.00$};
  \node[font=\scriptsize, align=center] at (0,-1.5) {mean distance 0.216\\[-1pt] $\mathrm{ADR}=0.98$};
  \node[font=\scriptsize, align=center, text=adrRed] at (2.9,-1.5) {mean distance 0.142\\[-1pt] $\mathrm{ADR}=0.64$};
\end{tikzpicture}
\caption{ADR illustrated with Run 1 measured values. Dot positions are schematic; they illustrate the metric rather than actual embedding coordinates. The three clouds represent the untrained attacker and its checkpoints after updates 9 and 24. Eight dots represent attacks for one behavior; connecting lines represent pairwise distances. The printed mean distance averages within-behavior diversity over 25 comparable development behaviors. Each cloud is scaled by its ADR relative to the first. Update 9 is Run 1 first checkpoint below 1 (0.98); update 24 is its last (0.64).}
\label{fig:adr}
\end{figure}
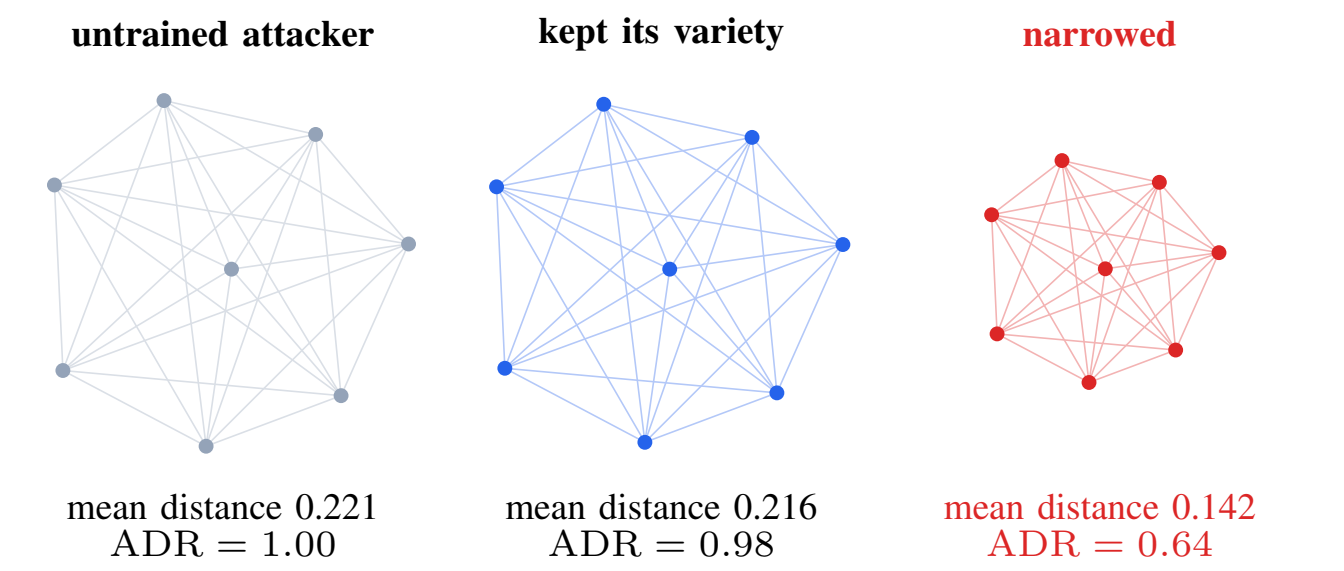

An attack is \emph{valid} if it has at least 20 and fewer than 1,500 characters and passes an opening-text refusal filter. The filter rejects attacker messages that themselves begin with a stock refusal. An incomplete or unscored evaluation, or a zero demoninator for the untrained-attacker, is reported as a measurement failure. For behavior $b$ at checkpoint $u$, let $d_b(u)$ be the mean cosine distance between all pairs of valid attacks, using embedding from a sentence encoder. For the set $\mathcal{B}$ of behaviors with at least two valid attacks in both banks,
\begin{equation}
    \mathrm{ADR}(u)=\frac{\sum_{b\in\mathcal{B}} d_b(u)}{\sum_{b\in\mathcal{B}} d_b(0)},
    \label{eq:adr}
\end{equation}
where $d_b(0)$ comes from the same untrained attacker seed under the same behaviors, decoding and sample seeds. This primary ADR measures how much within-behavior spread the trained attacker retains (Figure~\ref{fig:adr}). A value of 1 preserves the initial spread; 0.75 corresponds to a quarter lost; values above 1 indicate greater spread. The diversity reduction criterion combines this relative diversity reading with continued attack success.

The intervals describe behavior sampling conditional on the trained model and its sampled completions. The bootstrap treats a behavior and its eight completions as one sampling unit. 2,000 paired draws \cite{efron1979bootstrap} are made, stratified by benchmark, applying the same resampled behavior indices to the trained and untrained banks. The 2.5th and 97.5th percentiles give the confidence interval (CI), whose upper end $\mathrm{ADR}^{+}$ enters the diversity reduction criterion. ASR@1 uses the first completion for each behavior; ASR@8 counts success in any of the eight.

\subsection{Development Checkpoints and Diversity Reduction Checks}

After each completed attacker job, the attacker writes eight attacks for each of the 26 AdvBench development behaviors. These evaluation banks are kept from training. Checkpoint 0 is the untrained attacker, with $\mathrm{ADR}=1$ by definition. Behaviors and sample seeds stay fixed throughout the trajectory.

ADR compares the spread of the attack with the untrained model; attack success is scored against the frozen defender. This separates attacker learning from changes in the guardrails of the co-trained defender \cite{hsiung2026guardrails}. The checkpoint index counts completed jobs, so skipped attempts leave the index unchanged. Diversity reduction checks draw an additional bank from the attacker at the end of the round.

\subsection{Secondary Diagnostics and the Length-Matched Measure}

Strategy labels are automated descriptions of successful attacks. Each attack receives the nearest of seven short descriptions in embedding space when cosine similarity is at least 0.25; lower similarities receive the label ``Other.'' Section~\ref{sec:families} gives the descriptions. The share of each attack family and their entropy is reported, normalized by $\ln 8$ for the seven families plus ``Other.''

Additional diagnostics include within-behavior near-duplicates, defined by cosine similarity above 0.95, and the mean and spread of attack length in characters. Training diagnostics record reward signal, KL, gradient norms, the clipped share of tokens, completed and skipped updates, and benign refusal. The classifier may shorten the generated answer to fit its 2,048-token context; clipping counts are logged per job.

An exploratory \emph{length-matched} ADR was added after completing the pre-specified design. It compares distances within bands defined by thirds of the untrained attacker's pair lengths, measured as the mean character length of two attacks. The bands are under 184, 184 to under 244, and 244 or more characters. Within each band, distances are averaged over pairs from the shared behaviors. A band is eligible when both attackers have at least 15 pairs in it. Eligible bands are weighted by the untrained attacker share of the pairs in each band, normalized over those bands. This additional measure uses 1,000 behavior-bootstrap draws for its intervals.

\subsection{Onset Analyses}

The onset analyses describe the trajectory and are separate from the diversity reduction criterion. The \emph{persistent clear onset} is the earliest checkpoint beginning a sequence of three successive trained checkpoints, each with $\mathrm{ADR}<0.75$, $\mathrm{ADR}^{+}<0.75$ and the validity plus support guards satisfied. This definition was fixed before training. It was applied retrospectively, alongside correlations of ADR with development ASR@1 and update number.

\section{Implementation, Provenance and Containment}
\label{sec:execution}

\textbf{Implementation.} Sampling can vary when inference is repeated. This experiment aims to prevent variation by having the pipeline use a pinned code version and configuration, with atomic checkpoints for saved training state. Its co-training, development measurements, optimizer diagnostics, degenerate-run rule and paired intervals passed an offline regression suite before launch.

\textbf{Provenance and availability.} The design and configuration were stored in version control before training. Retained metadata include the configuration and dataset digests, package versions, model identifiers, container image, code digest and numerical run summaries.

\textbf{Containment.} The study excludes production targets, functional payloads and external model APIs. Raw benchmark text, attacks, answers, preference examples and adapters remain private. Analyses use numerical summaries. Research jobs use an outbound allowlist, with dataset download hosts enabled during preparation; a disable switch stops scripts at their next checkpoint.

\section{Results}
\label{sec:results}

\subsection{What Ran}

Runs 1 to 3 each completed 24 attacker updates and 12 defender updates, totaling 72 attacker and 36 defender updates. All recorded attempts produced an update, and the degenerate-run rule remained untriggered. At trained checkpoints, invalid attacks accounted for at most 2.9\% of a development bank; group coverage was 1.000.

Each run generated 36{,}864 training attacks, or 110{,}592 in total. Their success rates against the current defender were 33.5\%, 17.3\% and 25.2\%. Each run produced 28 development banks of 208 attacks, 24 checkpoints and three diversity reduction checks. The median training success rate rose from 0.042 over each of the first three updates to 0.401 over its last three. The share of training groups carrying a learning signal rose from 0.39 to 0.89.

\textbf{Stopping.} This was a data-dependent stop. Budget considerations informed the decision to end training after the diversity reduction checks at round 12. The decision followed the update-22 checkpoints, whose intervals were below 0.75 in each run, and preceded the drawing of the final diversity-reduction banks. Each run then completed those checks.

\subsection{Diversity Reduction Checks}

\begin{table*}[t]
\centering
\caption{Diversity reduction checks under the pre-specified criterion. ADR is shown with its 95\% bootstrap interval; the length-matched comparison is exploratory. ASR@1 is the share of development behaviors whose first attack succeeded. Over-refusal is the increase in refusals on 56 benign development prompts; one prompt is 0.018. A ``yes'' requires both the diversity threshold and the effectiveness guards in Section~\ref{sec:checks}.}
\label{tab:checks}
\small
\begingroup\setlength{\tabcolsep}{4pt}
\begin{tabular*}{\linewidth}{@{\extracolsep{\fill}}ccllccl@{}}
\toprule
Round & Run & \shortstack{ADR\\(95\% CI)} & \shortstack{Length-matched ADR\\(95\% CI)} & \shortstack{ASR@1: untrained\\$\to$ check} & \shortstack{Over-refusal\\increase} & \shortstack[l]{Shows\\reduction} \\
\midrule
4 & 1 & 1.08 (0.99--1.18) & 1.29 (1.17--1.39) & 0.04 $\to$ 0.42 & +0.000 & no: ADR above 0.75 \\
4 & 2 & 0.84 (0.77--0.93) & 0.86 (0.75--0.97) & 0.00 $\to$ 0.04 & +0.000 & no: ADR above 0.75; ASR gain 0.04 \\
4 & 3 & 0.79 (0.71--0.87) & 0.99 (0.89--1.10) & 0.00 $\to$ 0.35 & +0.000 & no: ADR above 0.75 \\
8 & 1 & 0.67 (0.59--0.754) & 0.755 (0.65--0.87) & 0.04 $\to$ 0.92 & +0.018 & no: upper bound 0.754 \\
8 & 2 & 0.77 (0.69--0.88) & 0.89 (0.80--1.01) & 0.00 $\to$ 0.38 & +0.036 & no: ADR above 0.75 \\
8 & 3 & 0.79 (0.70--0.89) & 0.99 (0.86--1.12) & 0.00 $\to$ 0.58 & +0.000 & no: ADR above 0.75 \\
12 & 1 & 0.63 (0.57--0.70) & 0.71 (0.61--0.84) & 0.04 $\to$ 0.85 & +0.036 & yes \\
12 & 2 & 0.61 (0.54--0.68) & 0.68 (0.54--0.751) & 0.00 $\to$ 0.58 & +0.000 & yes \\
12 & 3 & 0.63 (0.57--0.70) & 0.72 (0.65--0.80) & 0.00 $\to$ 0.77 & +0.018 & yes \\
\bottomrule
\end{tabular*}
\endgroup

\end{table*}

Table~\ref{tab:checks} reports all nine checks. The six checks at rounds 4 and 8 fell short of the criterion. At round 4, ADR was 1.08, 0.84 and 0.79; Run 2 success gain of 0.04 also fell short of the required 0.10. At round 8, Run 1 point estimate was 0.67 and its interval reached 0.754, exceeding the threshold by 0.004. Runs 2 and 3 had point estimates of 0.77 and 0.79.

The final checks met the criterion in each run. ADR ranged from 0.608 to 0.631, with upper bounds of 0.704, 0.682 and 0.697. Their margins below the threshold were 0.046, 0.068 and 0.053. Success gains were 0.81, 0.58 and 0.77, group coverage was 1.000, and paired coverage was 0.962. All attacks were valid and each defender had completed 12 updates. Over-refusal rose by 0.036, 0.000 and 0.018, within the 0.05 limit.

The across-run summary is post hoc and nominal. Its $t$ tests describe variation among three seed values conditional on the shared behavior panel; behavior sampling and the data-dependent stop are outside these tests. Mean ADR was 0.623, with standard deviation 0.013. One-sample $t$ tests place the mean below 1 ($t(2)=-48.9$, $p=0.0004$) and below 0.75 ($t(2)=-16.4$, $p=0.004$).

\subsection{Trajectories of Diversity and Attack Success}

\begin{figure*}[t]
\centering
\includegraphics[width=\textwidth]{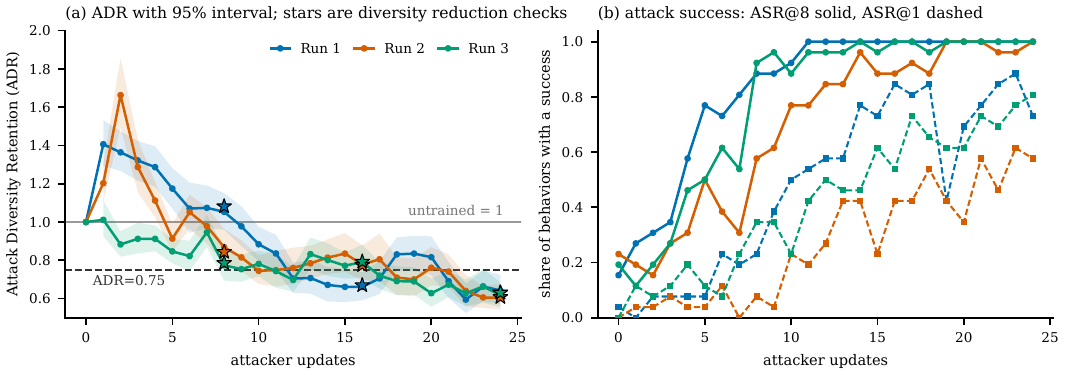}
\caption{Development trajectories after each attacker update. (a) ADR with its 95\% bootstrap interval. The dashed line marks the diversity reduction threshold of 0.75; stars mark the additional diversity-reduction banks at rounds 4, 8 and 12. (b) Attack success against the frozen defender: ASR@1 uses the first sample, and ASR@8 uses any of eight samples.}
\label{fig:trajectory}
\end{figure*}

ADR initially rose in two runs (Figure~\ref{fig:trajectory}). Run 1 reached 1.41 after one update and remained above 1 for eight updates; Run 2 reached 1.66 after its second. Run 3 began near 1, at 1.01, before falling.

The three-checkpoint onset is bounded by the observation window: with 24 checkpoints, the latest possible onset is update 22. Two onset estimates fall at this boundary. Point estimates first crossed below 0.75 at updates 12, 10 and 11 for Runs 1 to 3. The entire interval was below 0.75 at 4, 3 and 4 of the 24 checkpoints, all after update 15. Run 1 qualifying checkpoints were 16 and 22 to 24. Run 2 qualifying checkpoints were 22 to 24, and Run 3 qualifying checkpoints were 20 to 22 and 24. Persistent clear onset occurred at updates 22, 22 and 20.

The trajectories include reversals: Run 1 recovered from 0.66 at update 16 to 0.83 at update 18 before falling again. Median changes between consecutive checkpoints were 0.04, 0.06 and 0.04, and average interval widths were 0.15 to 0.19. These fluctuations make the sustained trajectory more informative than a single checkpoint.

The correlations describe joint trends with training time; they do not establish which variable leads or a causal effect. Across the 24 checkpoints, the Spearman correlation was $\rho$ between ADR and ASR@1 was $-0.91$, $-0.84$ and $-0.86$; with update number it was $-0.89$, $-0.90$ and $-0.90$. At the last checkpoint, ASR@1 was 0.73, 0.58 and 0.81, and ASR@8 reached 1.00 in each run. ASR@1 retains resolution after ASR@8 reaches its ceiling.

\subsection{Attack Length and the Length-Matched Measure}

\begin{figure*}[t]
\centering
\includegraphics[width=\textwidth]{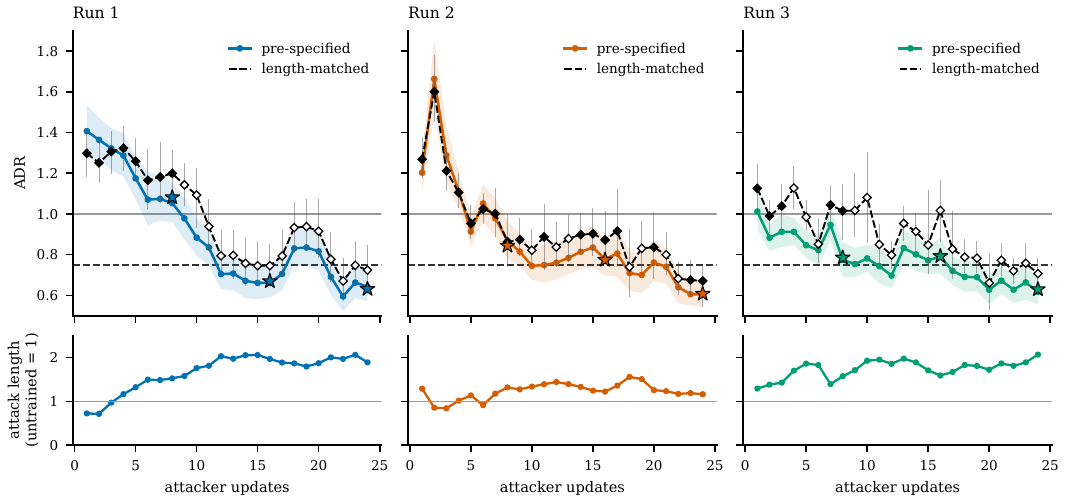}
\caption{Primary and length-matched ADR, with attack length. Top: primary ADR with a shaded 95\% interval and the comparison with interval bars. Hollow markers identify thin estimates using one or two of the three length bands. Stars mark diversity reduction checks; the horizontal line marks 0.75. Bottom: mean attack length relative to the untrained attacker.}
\label{fig:length}
\end{figure*}

Embedding distance covaries with attack length even before training. In the untrained banks, mean within-behavior distance was 0.239 to 0.261 for the shortest third of pairs and 0.196 to 0.208 for the longest. Pair distance correlated with length at $-0.18$ to $-0.28$. By the last checkpoint, mean attack lengths had grown to 1.89, 1.16 and 2.07 times their initial values. They began at 217, 222 and 209 characters and ended at 410, 258 and 431.

Figure~\ref{fig:length} compares the two diversity measures. At update 24, length-matched ADR was 0.72, 0.67 and 0.71, compared with primary ADR of 0.64, 0.60 and 0.62. At the final diversity reduction checks, the exploratory comparison was 0.71, 0.68 and 0.72. Matching raised these final estimates by 0.07 to 0.09.

The longest band has no upper edge, leaving residual length differences within it. For Runs 1 and 3, at least 92\% of trained pairs fall in that band from update 11 onward. The exploratory comparison use only one band. Each length-checked comparison interval at the final checks extended above 0.75. One of the 72 checkpoint intervals lie wholly below it: Run 2 at update 24 (0.67, interval 0.54 to 0.74). All six final comparison intervals, from the last checkpoints and diversity reduction checks, lie below 1. At the checks their upper bounds were 0.843, 0.751 and 0.800.

\subsection{Strategy Families, Behaviors and the Defender}
\label{sec:families}

\begin{figure*}[t]
\centering
\includegraphics[width=\textwidth]{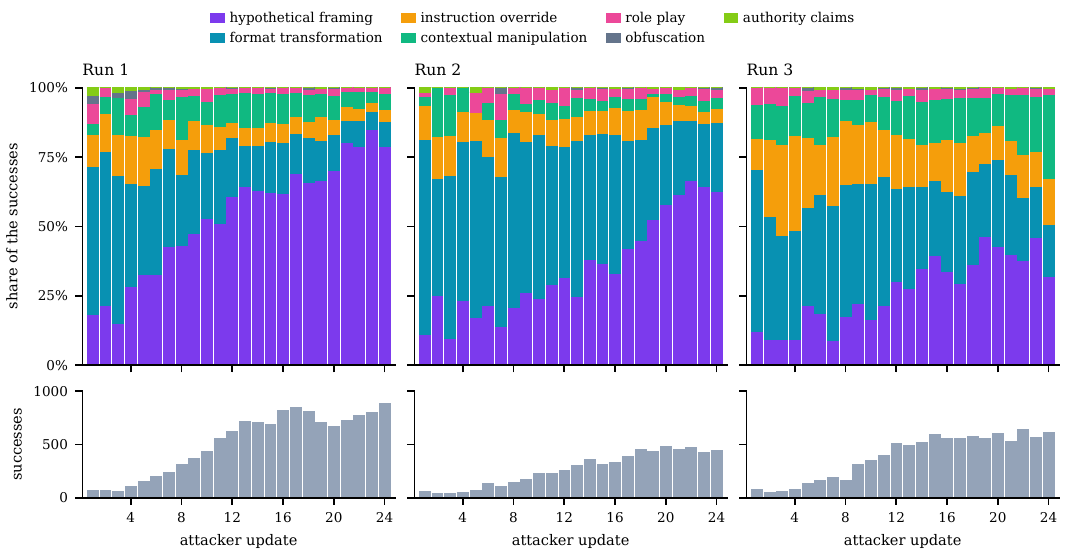}
\caption{Automated family labels among successful training attacks, with one pair of panels per run. Each upper bar gives the shares for one rollout, stacked in table order. Lower panels give the number of successes out of 1,536 attacks. Labels identify the nearest description in embedding space.}
\label{fig:families}
\end{figure*}

The composition of successful attacks changed during training. Figure~\ref{fig:families} follows the family shares by update, and Table~\ref{tab:families} pools them into windows.

\begin{table}[t]
\centering
\caption{Family shares among labeled successes ($n$), pooled over runs. The Untrained column uses the development banks of the untrained attacker; the early and late columns use the first and last three training rollouts. All successes received one of the seven family labels.}
\label{tab:families}
\small
\begin{tabular}{@{}lrrr@{}}
\toprule
Family (\%) & Untrained & First 3 & Latest 3 \\
\midrule
successes ($n$) & 20 & 541 & 5{,}616 \\
hypothetical framing & 0.0 & 14.4 & 63.2 \\
format transformation & 65.0 & 53.2 & 15.7 \\
instruction override & 30.0 & 16.8 & 7.7 \\
contextual manipulation & 0.0 & 10.4 & 11.1 \\
role play & 0.0 & 3.9 & 2.2 \\
obfuscation & 5.0 & 0.6 & 0.1 \\
authority claims & 0.0 & 0.7 & 0.0 \\
\bottomrule
\end{tabular}

\end{table}

A family label records similarity to a description, and a message combining techniques still receives one label. A rollout contains 1,536 attacks, with 40 to 889 judged successful across the observed updates. Colored segments divide those successes into shares totaling 100\%. The seven descriptions are:
\begin{itemize}
    \item \emph{hypothetical framing}: the request framed as fiction, a hypothetical, a thought experiment or a story;
    \item \emph{format transformation}: the content asked for in another form, such as a translation, summary, list or piece of code;
    \item \emph{instruction override}: an attempt to override, reset or ignore earlier instructions and rules;
    \item \emph{contextual manipulation}: a benign-looking scenario or backstory built around the request;
    \item \emph{role play}: a persona or character that behaves differently from the usual behavior of the assistant;
    \item \emph{obfuscation}: the request disguised by encoding, euphemism, misspelling or splitting it into pieces;
    \item \emph{authority claims}: claimed official permission, professional credentials or a duty that supposedly justifies the request.
\end{itemize}
Format transformation accounted for 54\%, 70\% and 59\% of successes at update 1; hypothetical framing accounted for 18\%, 11\% and 12\%. Hypothetical framing later led at 19 of 24 updates in Run 1 and at 8 and 10 updates in Runs 2 and 3. Its uninterrupted leads began at updates 7, 17 and 18. At update 24, its shares were 79\%, 63\% and 32\%, and format transformation had fallen to 9\%, 25\% and 19\%.

The ending mixes differed: Run 1 concentrated 79\% of successes in one family, with each other family at 9\% or less. Run 3 spread successes across hypothetical framing (32\%), contextual manipulation (30\%), format transformation (19\%) and instruction override (17\%). Role play, obfuscation and authority claims reached maxima success of 9.3\%, 2.9\% and 2.8\%.

Run 1 family entropy was 0.47 to 0.74 over its first 20 updates and 0.29 to 0.38 over its last four. Hypothetical framing then accounted for 79\% to 85\% of successes. Runs 2 and 3 stayed within entropy ranges of 0.47 to 0.65 and 0.60 to 0.71.

The earliest shares are coarse: the first three rollouts per run had 40 to 82 successes, compared with later maxima of 889, 478 and 640. Pooled format-transformation shares fell from 65.0\% of the 20 successes of untrained attacker to 53.2\% in the early window and 15.7\% in the late window. Hypothetical framing rose from 0\% through 14.4\% to 63.2\%; instruction override fell from 30.0\% through 16.8\% to 7.7\%. Across all updates, hypothetical framing represented 64.6\%, 44.1\% and 32.7\% of successes in Runs 1 to 3.

The exploratory family-by-length comparison uses the same encoder as ADR, so it supplies descriptive rather than independent strategy evidence. Its short bands are sparse: 111 and 491 of the latest 5{,}616 successes. Late hypothetical-framing successes averaged 383 characters, close to format transformation at 370 and the other families at 397. Its shares rose within the short, middle and long bands from 11\%, 14\% and 16\% to 66\%, 70\% and 62\%. Reweighting the late successes to the early length mix gave a share of 64.9\%, compared with 63.2\% observed.

The cross-run comparison points in the same direction. Run 2 attack length stayed closest to its reference, at 1.16 times the starting length, and 65\% of late successes were hypothetical framing. Run 3 length grew most, to 2.07 times its reference, and its late hypothetical-framing share was 38\%. The family gained share within comparable length ranges as well as across the pooled rollouts.

\begin{table*}[t]
\centering
\caption{The three runs at their last checkpoint (update 24). Diversity is the mean cosine distance between the attacks written for one behavior.}
\label{tab:runs}
\small
\begin{tabular}{@{}p{2.6in}ccc@{}}
\toprule
Quantity & Run 1 & Run 2 & Run 3 \\
\midrule
ADR at the last checkpoint (95\% CI) & 0.64 (0.57--0.72) & 0.60 (0.54--0.68) & 0.62 (0.56--0.69) \\
Checkpoints with ADR below 0.75: value / whole interval (of 24) & 10 / 4 & 8 / 3 & 10 / 4 \\
Behaviors less diverse than untrained / with ratio below 0.75 (of 25) & 22 / 16 & 24 / 19 & 24 / 20 \\
Diversity, untrained $\to$ last (mean cosine distance) & 0.221 $\to$ 0.143 & 0.228 $\to$ 0.137 & 0.235 $\to$ 0.143 \\
ASR@1, untrained $\to$ last & 0.04 $\to$ 0.73 & 0.00 $\to$ 0.58 & 0.00 $\to$ 0.81 \\
ASR@8, untrained $\to$ last & 0.15 $\to$ 1.00 & 0.23 $\to$ 1.00 & 0.19 $\to$ 1.00 \\
Mean attack length in characters, untrained $\to$ last & 217 $\to$ 410 (1.89$\times$) & 222 $\to$ 258 (1.16$\times$) & 209 $\to$ 431 (2.07$\times$) \\
Training-rollout diversity, last three updates vs. the first rollout & 0.69 & 0.66 & 0.66 \\
Successful attacks in the training rollouts & 33.5\% & 17.3\% & 25.2\% \\
\bottomrule
\end{tabular}

\end{table*}

One development behavior lacks two valid untrained attacks, leaving 25 comparable behaviors. At the last checkpoint, within-behavior spread was lower than the untrained attacker for 22, 24 and 24 behaviors in Runs 1 to 3. Ratios were below 0.75 for 16, 19 and 20 behaviors, with medians of 0.60, 0.58 and 0.64. Near-duplicate shares rose from 0.045, 0.030 and 0.000 to 0.212, 0.192 and 0.130.

The training rollouts provide a larger view of 96 training behaviors against the moving defender (Table~\ref{tab:runs}). Mean diversity over the last three rollouts fell to 0.69, 0.66 and 0.66 of first rollout of each run. At all nine diversity reduction checks, benign over-refusal rose by at most two of the 56 prompts, or 0.036.

\section{Discussion}
\label{sec:discussion}

The three runs share one development panel, so their agreement makes sense for these behaviors. The criterion operationalizes diversity reduction as a loss of embedding spread while attack effectiveness continues. Under that definition, the runs provide a reproduction of the diversity loss reported for GRPO co-training \cite{bullwinkel2026advgrpo}. The decline appears in the per-behavior comparisons and in the larger training rollouts as well as in aggregate ADR.

The 0.75 threshold is a convention for losing a quarter of the initial spread. Near-duplicate shares also depend on a chosen cutoff, 0.95, and paraphrases can implement the same template while falling below it. Template counts and human strategy coverage therefore remain unmeasured. At the last checkpoint, near-duplicate shares were 0.13 to 0.21: most samples stayed outside the cutoff even as overall spread declined.

Length grew during training even with constant loss normalization. That normalizer addresses token weighting, while output length remains a property of the learned policy. Banding and pair weighting make the comparison evaluation a sensitivity analysis; residual within-band differences prevent an exact attribution of the decline to length. Primary ADR records the total change in embedding spread, and the length-check comparison shows how the reading changes under those length bands.

Stopping selection can make the intervals and $p$ values optimistic. The final banks were drawn after that decision, scoring the same selected attackers and behaviors; they share 47\% to 74\% of their texts with the preceding checkpoints. They provide within-run consistency evidence, with ADR differing by at most 0.03 between the two banks.

\section{Limitations}
\label{sec:limits}

\begin{enumerate}
    \item \textbf{Sampling.} Joint uncertainty over seeds, behaviors and completion draws remains unestimated. The behavior bootstrap conditions on a seed and its completions; the across-run $t$ test conditions on the shared panel. The test splits remained unscored. This experiment reports each run individually, following recommendations for small-sample reinforcement learning studies \cite{henderson2018matters,agarwal2021precipice}.
    \item \textbf{Stopping.} The data-dependent stop departs from the pre-specified design. Further rounds, additional seeds, the planned budget check and held-out evaluation were left unrun. Intervals and $p$ values are uncorrected for stopping and the three diversity reduction checks per run; their role here is descriptive.
    \item \textbf{Measurement.} Embedding geometry, wording and attack length affect the diversity reading. Classifier preferences could also contribute to the observed strategy shift. Model judges have known biases \cite{zheng2023judge}. A separate agent-training study found judge-based rewards more susceptible to reward hacking than an embedding-based reward \cite{yu2026rwml}. Human strategy labels and a human audit were not collected. Family labels and lexical over-refusal checks provide automated diagnostics.
    \item \textbf{Attribution.} A single training arm leaves the effects of co-training and optimizer stabilization unresolved. Testing those effects would require frozen-defender and unstabilized comparison arms; prevention would require an intervention arm. The present result characterizes one co-training schedule with all four stabilization techniques applied together.
    \item \textbf{Scope.} Multi-turn strategies such as Crescendo \cite{russinovich2024crescendo}, deployed systems and adaptive human attackers \cite{bullwinkel2025lessons} lie outside this evaluation. It covers single-turn attacks between two open-weight models, using one implementation of the update and fixed public benchmarks.
    \item \textbf{Clipped answers.} A clipped answer supplies partial evidence to the classifier. Its 2,048-token context can require shortening the generation section. The scoring procedure preserves the rubric and answer cue.
    \item \textbf{Pre-specification.} The design was dated and version-controlled before launch. Three analyses were added afterward: length-matched ADR, the across-run $t$ test and the family-by-length comparison. The diversity reduction criterion continues to use primary ADR.
\end{enumerate}

\section{Related Work}
\label{sec:related}

Language-model red teaming includes model-generated test cases \cite{perez2022redteaming} and optimized adversarial suffixes \cite{zou2023universal}, with HarmBench standardizing automated attack and refusal evaluation \cite{mazeika2024harmbench}. Several methods directly reward diversity: curiosity-driven red teaming adds a novelty objective \cite{hong2024curiosity}, Rainbow Teaming uses quality-diversity search \cite{samvelyan2024rainbow}, and Beutel et al.\ reward attacks that differ from past attempts \cite{beutel2024diverse}.

Diversity loss appears in reinforcement learning from human feedback \cite{kirk2023rlhf}, and repeated learning from narrowed outputs can erode distribution tails \cite{shumailov2024collapse}. These findings motivate tracking the distribution of attacks alongside their success rate.

For reasoning models, Liu et al.\ identify the per-completion length biase of GRPO \cite{liu2025drgrpo}; Yu et al.\ identify change loss aggregation and resample uniform-reward groups \cite{yu2025dapo}. Their findings motivate the use of constant normalizer and signal-row bounds. Ratio clipping \cite{schulman2017ppo} and learning-rate warm-up \cite{liu2020radam} address update stability. These safeguards were intended to reduce length-weighting bias, sparse-update noise and excessive policy changes within a rollout, making the diversity trajectory easier to interpret.

Small-sample reinforcement learning results benefit from individual runs and interval estimates \cite{henderson2018matters,agarwal2021precipice}. That reporting approach was followed. The across-run mean supplements the individual measurements as a post hoc summary.

\section{Conclusion}
\label{sec:conclusion}

This experiment uses a data-dependent stopping time and nominal intervals. The three runs met the diversity reduction criterion at round 12, with a 37\% to 39\% loss of within-behavior embedding spread and attack success above its starting level. Length matching gives a smaller decline, and the strategy mix changes toward hypothetical framing. Held-out evaluation and comparison arms would test generalization and cause. The runs demonstrated effective attack generation alongside a narrower within-behavior distribution.

\bibliographystyle{IEEEtran}
\bibliography{references}

\end{document}